\documentstyle[12pt]{article}
\newlength{\dinwidth}
\newlength{\dinmargin}
\newcommand{\ba}{\begin{array}}
\newcommand{\ea}{\end{array}}
\newcommand{\be}{\begin{equation}}
\newcommand{\ee}{\end{equation}}
\newcommand{\bea}{\begin{eqnarray}}
\newcommand{\eea}{\end{eqnarray}}

\def\bee{\begin{eqnarray}}
\def\eee{\end{eqnarray}}
\def\be{\begin{equation}}
\def\ee{\end{equation}}

\begin{document}
\thispagestyle{empty}
\addtocounter{page}{-1}
\begin{flushright}
SNUTP 98-119\\
{\tt hep-th/9811008}\\
\end{flushright}
\vspace*{1.3cm}
\centerline{\Large \bf Holographic Principle and String 
Cosmology~\footnote{
Work supported in part by KOSEF Interdisciplinary Research Grant and 
SRC-Program, KRF International Collaboration Grant, 
Ministry of Education Grant BSRI 98-2418 and 98-2425, UOS Academic
Research Program, SNU Faculty Research Grant, and The Korea Foundation for 
Advanced Studies Faculty Fellowship.}}
\vspace*{1.2cm} \centerline{\large \bf Dongsu Bak${}^1$ and Soo-Jong Rey${}^2$}
\vspace*{0.8cm}
\centerline{\large\it Physics Department, 
University of Seoul, Seoul 130-743 KOREA${}^1$}
\vskip0.3cm
\centerline{\large\it Physics Department, Seoul National University,
Seoul 151-742 KOREA${}^2$}
\vskip0.4cm
\centerline{\tt dsbak@mach.uos.ac.kr, \quad sjrey@gravity.snu.ac.kr}
\vspace*{1.5cm}
\centerline{\Large\bf abstract}
\vspace*{0.5cm}
String cosmology is revisited from cosmological viewpoint of holographic 
principle put forward by `t Hooft, and by Fischler and Susskind. 
It is shown that the holography requires existence of a `graceful exit' 
mechanism, which renders the Universe nonsingular by connecting pre- and 
post-big bang phases smoothly. 
It is proven that flat Universe is consistent with the holography only if it 
starts with an absolutely cold and vacuous state and particle entropy is 
produced during the `graceful exit' period. Open Universe can always 
satisfy the holography no matter what initial state of the Universe is.
\vspace*{1.1cm}

\baselineskip=18pt
\newpage

\setcounter{equation}{0}

An emerging new paradigm in quantum gravity is the {\sl holographic principle},
first put forward by `t Hooft~\cite{thooft} 
in the context of black hole physics and more
recently extended by Susskind~\cite{susskind} 
to string theory. The principle demands that 
the degrees of freedom of a spatial region reside not in the bulk but only at
the boundary of the region and that the number of degrees of freedom per
Planck area is no greater than unity. 
 
Actually, basic ingredients leading to 
the holographic behavior seem to be quite generic in any gravitational (or 
diffeomorphism invariant) system that the principle should be equally 
applicable to a cosmological context. In fact, recently, Fischler and 
Susskind~\cite{fischlersusskind} 
have put forward a cosmological version of the holographic principle: at any
time during cosmological evolution, the gravitational entropy within a closed
surface should be always larger than the particle entropy that passes through
the past light-cone of that surface. Applied to the standard big-bang 
cosmology, Fischler and Susskind have found that only open or flat universe
but not closed one is compatible with the cosmological holographic principle,
provided one makes certain assumptions on the initial big-bang singularity.

In string theory, as first shown by Veneziano~\cite{veneziano},
cosmological evolution of
the Universe is drastically different from that in the standard big-bang
cosmology. In addition to post-big bang phase with an initial singularity, 
in string theory, a super-inflationary pre-big bang phase with a final 
singularity is present. Most interestingly, the pre-big bang phase may be 
juxtaposed with the post-big bang branch smoothly (referred as `branch change'
via `graceful exit'~\cite{gasperiniveneziano})
so that the singularities of both phases are simultaneously 
smoothened out. This opens up an exciting possibility in string theory of 
realizing an eternally evolving, non-singular and inflationary universe. 
While physical detail
of the `branch change', where curvature and quantum effects
are most pronounced, is yet to be understood thoroughly, 
recent studies have indicated that the
`branch change' could be triggered by higher-curvature~\cite{curvature},
massive string state~\cite{massive}, and quantum~\cite{quantumloop} back 
reactions. 

In this paper, we put the string cosmology to a test of 
the cosmological holographic principle, assuming homogeneity 
and isotropy and applying the holography in string frame to the 
Friedmann-Robertson-Walker (FRW) Universe with flat or open spatial 
curvature~\footnote{Closed Universe will not be considered here since, as 
Veneziano~\cite{veneziano} has emphasized, the very presence of initial and
final singularities contradicts with 
the basic postulates of the non-singular string cosmology.}. Summarizing the 
results, we find that flat Universe is ruled out by the principle {\sl unless}
the Universe is absolutely cold and vacuous throughout 
the pre-big bang phase. We 
also find that open universe can be consistent with the holographic 
principle, provided there exists a suitable mechanism of the `graceful exit'.
While admittedly preliminary, the most compatible string cosmology
with the holographic principle singles out the open Universe. 
Intriguingly, such an open Universe appears to be favored by recent 
measurement of the redshift-to-distance relation
and theory of the large-scale structure formation~\cite{berkeley}.
\vskip0.5cm
\sl Cosmological Holographic Principle: \rm
We begin with recapitulating the cosmological holographic principle. While the
principle is applicable to any cosmological context, in this paper, we will be 
mostly applying the principle for semi-classical string theory. 

At leading order in $\alpha'$- and string-loop expansions, the low-energy
string effective action in $(d+1)$-dimensional spacetime is given by
\be
S = {V_{9-d} \over l_s^{d-1}} \int d^{d+1} x \, \sqrt{-g}
e^{- 2 \phi} \left[ - R - 4 (\nabla \phi)^2 + {\cal L}_{\rm M} \right],
\label{lagrangian}
\ee
where $V_{9-d}, l_s, \phi$ denote the volume of the $(9-d)$-dimensional 
compactified space, the string length scale and the dilaton field respectively.
The Lagrangian ${\cal L}_{\rm M}$ summarizes matter contribution from 
(Neveu-Schwarz)$\times$(Neveu-Schwarz) sector. The $(d+1)$-dimensional 
effective Planck length scale, $l_P(t)$, depends on the dilaton field 
by~\footnote{Hereafter, we will take $l_s = 1$.}
\be
l_P(t) = l_s e^{2 \phi \over d-1} \, .
\ee
We assume that the universe is spatially homogeneous and isotropic and hence
is described by the FRW metric in string frame:
\be
ds^2_{\rm string} = dt^2 - a^2(t) \Big[ {dr^2 \over 1 - k r^2}
+ r^2 d \Omega^2_{d-1} \Big] \, ,
\label{metric}
\ee
where $k = 0, -1, +1$ correspond to a flat, open or closed Universe 
respectively. We will also assume that the NS-NS matter in ${\cal L}_{\rm M}$
is described by an ideal gas with equation of state $p = \gamma \rho$ for 
some constant $\gamma$. The equations of motion derived from 
Eq.(\ref{lagrangian}) read
\bee
{1 \over 2} d(d-1) \left( H^2 + {k \over a^2} \right) + 2 \dot \phi^2 
- 2d H \dot \phi &=& e^{2 \phi} \, \rho   
\nonumber \\
d H^2 + (d-1) {k \over a^2} + \dot H - 2 H \dot \phi &=& e^{2 \phi} p
\nonumber \\
4 \dot \phi^2 - 4d H \dot \phi + 2 d \dot H - 2 \ddot \phi 
+ d(d-1) \left( H^2 + {k \over a^2} \right) &=& 0 \, , 
\label{equationsofmotion}
\eee
where $H = \dot a / a$ is the Hubble parameter in string frame, and
$\rho$ and $p$ are the effective matter density and pressure. The above set
of equations implies that the matter density and pressure are related by  
\be
\dot \rho + d H (\rho + p) = 0 \, .
\label{bianchi}
\ee

The cosmological holographic principle is a direct extension of `t Hooft's
holography~\cite{thooft}. 
Consider a spatial region of coordinate size $R$, whose {\sl past}
light-cone converges to a point. The encolsed four-volume is a well-posed
physical system with a given light-cone data. 
Assuming that `t Hooft's holography is applicable to this physical region, 
the particle entropy inside the three-volume should be bounded by the 
gravitational entropy measured by the physical boundary area for all such 
radius $R$ and time $t$. That is, 
denoting particle entropy density in comoving coordinate $\sigma$, which is
constant by virtue of Eq.(\ref{bianchi}), comoving three-volume ${\cal V}(R)$,
and coordinate boundary area ${\cal A}(R)$, the holography condition is  
\be
\sigma {\cal V}(R) \le {a^{d-1}(t) {\cal A} (R) \over l_P^{d-1} (t) } . 
\label{thooft}
\ee
We will call the above as the `t Hooft holography condition.
Note that all the geometric quantities involved (area, volume, four-volume) are 
notions rather independent of cosmology. 

For the case of the standard cosmology, where an initial singularity is always
present, Fischler and Susskind have shown that the cosmological holographic
principle actually 
leads to a {\sl single} condition involving  the {\sl particle horizon} 
$R_H(t)$ measured
in comoving coordinates~\footnote{The particle horizon is independent of the
choice between the string and the Einstein frame.} 
$R_H(t) = \int_{t_I}^t d \tau / a(\tau)$,
where $t_I$ denotes the initial moment of the Universe. For a spherical 
region inside $R_H(t)$, whose surface area is $A_H(t)$ and enclosed volume
is $V_H(t)$ (both measured in comoving coordinates), the cosmological 
holographic principle requires that an inequality between the particle entropy
and the gravitational entropy
\be
\sigma V_H (t) \, < \, {a^{d-1}(t) A_H(t) \over l_P^{d-1}(t) } \, ,
\label{fsbound}
\ee 
should hold for all time $t$, both past and future. 

We will now test the holography on flat and open
string Universe separately, first with the criterion Eq.(\ref{fsbound}). 
We will then show that the more general, `t Hooft criterion Eq.(\ref{thooft}) 
yields essentially the same result. 

\vskip0.5cm
{\sl Flat Universe}: For flat Universe, the holography condition 
Eq.(\ref{fsbound}) may be rewritten as~\footnote{ The $\tilde{a}(t)$ turns
out to coincide with the sacle factor in Einstein frame.} 
\be
{R_H(t) \over \tilde{a}^{d-1}(t)} \, < \, {d \over \sigma},
\quad \left( \tilde{a}(t) \equiv e^{-2 {\phi \over d -1}} a(t) \right).
\label{flatbound}
\ee
As will be shown shortly, the pre-big bang phase of the flat Universe ranges 
$ -\infty < t < 0$ ( viz. $t_I \sim - \infty$), over which
$a(t) \sim a_0 (-t)^A$ and $\tilde{a}(t) \sim \tilde{a}_0 (-t)^B$. 
Importantly, the exponent $A$ is bounded above, $A < 1$. Hence, at any time
$t$, 
\be
R_H(t) = {(-t_I)^{1-A} - (-t)^{1-A} \over (1 - A) a_0} \sim \infty \, ;
\quad
\tilde{a}(t) \sim {\rm finite}
\ee
and the inequality Eq.(\ref{flatbound}) cannot be satisfied at any $t$. 
Apparently, the holography alone rules out the flat Universe in string 
cosmology. An interesting exception is if the flat Universe is absolutely 
cold and vacuous, viz. $\sigma = 0$. 
Then, the holography condition Eq.(9) is trivially satisfied 
until the `graceful exit' epoch, during which particle production will produce 
an entropy for all subsequent epoch. We thus conclude that the holography 
requires the flat string Universe should start not with matter but only with 
a cold and vacuous pre-big bang phase.

To verify the afore-mentioned behavior, we now solve
Eqs.(\ref{equationsofmotion}). Analytic solution of the pre-big bang phase 
in general $d$-spatial dimensions is given by:
\bee 
a(t) &=& a_0 (-t)^{2 \gamma / ( 1 + d \gamma^2)}
\nonumber \\
\phi(t) &=& \phi_0 + {d\gamma - 1 \over 1 + d \gamma^2} \ln (-t)
\eee
over the range $ -\infty < t < 0$. Here, $a_0, \phi_0$ are overall constant
factors and the coordinate time $t$ is measured in string unit. 
If the Universe is dominated by the dilaton, 
\bee
a(t) &=& a_0 (-t)^{-{1 / \sqrt{d}}}
\nonumber \\
\phi(t) &=& \phi_0 - {1 + \sqrt{d} \over 2} \ln (-t).
\label{flatdilatondominated}
\eee
We observe that, as claimed,  
the initial moment $t_I$ extends to $-\infty$ and the scale 
factor $a(t)$ behaves as $a_0 (-t)^A$ with $A < 1$.  
\vskip0.5cm
{\sl Open Universe:}
For an open Universe, we find it more convenient to rewrite
Eq.(\ref{metric})
\be
ds^2_{\rm string} = a^2(\eta) \left[ d \eta^2 - (d \chi^2 + 
\sinh^2 \chi d \Omega^2_{d-1} ) \right] \, , 
\label{conformalmetric}
\ee
where we have introduced the conformal time 
\be
\eta \equiv \int_0^t {d \tau \over a(\tau)} \, .
\ee
The particle horizon $R_H(\eta)$ measured in conformal coordinates is simply 
$R_H(\eta) = \eta - \eta_I$, where
$\eta_I$ denotes the initial moment of the Universe. For the pre-big bang 
phase, $ -\infty < \eta < 0$. Hence, $R_H (\eta) \sim \infty$ for all time 
and, using this fact, 
\be
{V_H (\eta) \over A_H (\eta)} 
= {\int_0^{R_H(\eta)} d \chi \sinh^{d-1} \chi \over \sinh^{d-1} R_H}
= {1 \over d -1}.
\label{openratio}
\ee
This is a remarkable relation in that the volume to area ratio in comoving 
coordinates stays constant, in sharp contrast to the flat Universe case. 
It is equally remarkable in that the {\sl spatial} volume-to-area relation 
Eq.(\ref{openratio}) of the open Universe is strikingly similar to the 
{\sl spacetime} volume-to-area relation \cite{susskindwitten} of the anti-de 
Sitter space, which also stays constant.

The holography condition Eq.(\ref{fsbound}) is now rephrased to
\be
{1 \over \tilde{a}^{d-1}(\eta)} \, < \, {d-1 \over \sigma}.
\label{openbound}
\ee
Thus, potentially the most dangerous epoch is when ${\tilde a} 
\rightarrow 0$. It turns out, during this epoch, 
$\tilde{a}(\eta) \sim \tilde{a}_0 
(-\eta)^{\tilde B}$, as will be shown shortly.
For $\gamma < 1$, $\tilde{B} > 0$ and $\tilde{a}(\eta)$
decreases monotonically. This appears to imply that the holography is violated 
near the final singularity of the pre-big bang phase, $\eta \sim
0$. However, if the pre-big bang phase makes a graceful exit to the post-big
bang phase smoothly, then $\tilde{a}(\eta)$ bounces back before the Universe
reaches the final singularity $\eta \sim 0$ and the holography
can be satisfied. If $\gamma > 1$, $\tilde{B}$ may be negative and 
$\tilde{a}(\eta)$
may increase monotonically. In this case, the holography is bound to fail 
incurably at the initial moment $\eta \sim \eta_I = - \infty$. Intriguingly,
$\gamma > 1$, which is inconsistent with special relativity, is also 
 inconsistent with the holographic principle. We thus conclude that open
Universe can be compatible with the cosmological holographic principle 
{\sl provided} $\gamma < 1 $ and the pre-big bang phase makes successfully a 
graceful exit to the post-big bang phase. 

We now verify that $\tilde{a}(\eta) \sim \tilde{a}_0 (-\eta)^{\tilde B}$.
In terms of the conformal time, the Einstein equations Eqs.(\ref{equationsofmotion}) can be presented by 
\bee
(\tilde{a}')^2 &=& - k \tilde{a}^2 + {2 \tilde{a}^2 \over d(d-1)}
\left[ \rho e^{{2(d+1) \over d-1} \phi} + {2 \over d-1} {(\phi')^2 
\over \tilde{a}^2} \right]
\label{confomaleq}\\  
\phi'' &+& (d-1) {\tilde{a}' \over \tilde{a}} \phi' = 
{1 \over 2} (dp - \rho) (\tilde{a})^2 e^{{2 (d+1) \over d-1} \phi} \, ,
\label{conformalequationofmotion}
\eee
where the primes denote derivatives with respect to the conformal time $\eta$.
From Eq.(\ref{bianchi}) and the equation of state, we further find
\be
\rho = \rho_0 \tilde{a}^{-d(1 + \gamma)} e^{{2d (1 + \gamma) \over d-1} \phi}
\, .
\ee
For the holography bound Eq.(\ref{openbound}), potential
violation of the holography occurs whenever $\tilde{a} \sim 0$ so that, from 
Eqs.(\ref{confomaleq})-(\ref{conformalequationofmotion}),
the spatial curvature term is sub-dominant over the remaining terms. 
During this epoch,
\bee
\tilde{a}(\eta) &\sim& \tilde{a}_0 (-\eta)^{2 (1 - \gamma) \over
(d-1) (d\gamma^2 - 2\gamma + 1)}
\nonumber \\
\phi(\eta) &\sim& \phi_0 + {d\gamma - 1 \over d\gamma^2 - 2 \gamma + 1}
\ln (-\eta).
\eee
If the Universe is dilaton dominated,
\bee
\tilde{a}(\eta) &\sim & \tilde{a}_0 (-\eta)^{1 \over d - 1}
\nonumber \\
\phi(\eta) &\sim& \phi_0 \mp {\sqrt{d} \over 2} \ln (-\eta).
\eee
We then recognize immediately the claimed behavior of the scale factor:
$\tilde{a}(\eta) \sim 
\tilde{a}_0 (-\eta)^{\tilde B}$, where ${\tilde B} > 0$ if $\gamma <1$ and
${\tilde B}$ may be negative if $\gamma > 1$.
Since the right hand side of Eq.~(\ref{confomaleq}) 
(with $k=-1$) is 
positive definite 
for $-\infty < \eta < 0$, the signature of   
$\tilde{a}' <0$ remains the same all the time. Consequently, 
the scale factor $\tilde{a}$ is monotonically 
decreasing for $\gamma <1$, whereas it may be 
monotonically increasing 
for $\gamma>1$.
 \vskip0.5cm 
{\sl The `t Hooft's Holography Condition:}
Actually, the conclusion drawn above based on the more specific, 
Fischler-Susskind condition also holds even if we apply the 
`t Hooft condition. The `t Hooft holography condition Eq.(\ref{thooft}) 
is essentially the same as 
Eqs.(\ref{flatbound},\ref{openratio},\ref{openbound}), 
except now that $R_H$ should be interpreted as
the comoving size $R$ of a fiducial region, namely, 
$R /\tilde{a}^{d-1}(t) < d/\sigma$ for the flat Universe and
$ ( \sigma\int_0^{R} d \chi \sinh^{d-1} \chi )/( 
\tilde{a}^{d-1}(t) \sinh^{d-1} R )< 1$ for the open Universe.
According to the `t Hooft's condition, the holography should then hold for 
{\sl all} choices of comoving size $R$ and time, as there is 
{\sl no} beginning of the Universe in string cosmology. 
In particular, it should be so for the comoving size $R$ arbitrarily large. 
This immediately leads basically to the same conclusion as the 
Fischler-Susskind holography condition, even though the reasoning
involved is completely different. 

\vskip0.5cm
{\sl Consideration of Post-Big Bang Phase}:
It is also of interest to test the holography to the post-big bang phase.
We only need to recall that the post-big bang phase is related to the
pre-big bang one by simultaneous application of two symmetry transformations:
scale-factor duality $a(t) \rightarrow 1/a(t)$, $\phi(t) \rightarrow \phi(t) 
- d \ln a(t)$, and time-reversal symmetry $-t \rightarrow +t$.

The flat Universe evolves as $a(t) \sim a_0 t^C$ with $C < 1$ during the 
post-big bang phase, $0 < t < + \infty$. We thus find that 
\be
R_H(t) \sim {\rm constant} \cdot t^{1 - C}; \quad
 {V_H(t) \over \tilde{a}^{d-1}(t) A_H(t)} 
\sim {\rm constant} \cdot t^{{(d \gamma^2 - 1) \over (d \gamma^2 + 1)}}
\, .
\label{postflatcond}
\ee
Hence, for the post-big bang matter with $ \vert \gamma \vert < 1/ \sqrt{d}$
, the holography condition is satisfied for all later time {\sl provided} 
the condition is satisfied at the Planck epoch. 
The condition at the Planck epoch puts a severe constraint on the 
`graceful exit' mechanism since, as we have shown earlier, the Universe should
be cold and vacuous during the pre-big bang phase and all the entropy should
be created during the `graceful exit' epoch, whose size is now constrained
by the holography Eq.(\ref{postflatcond}).
Note also that, after the `graceful exit' is taken into account, `effective'
particle horizon $R_H^{\rm eff}(t) = \int_{-\infty}^t d \tau / a(\tau)$
becomes infinite for {\sl all} time.

For open Universe, during the post-big bang phase $0 < \eta < + \infty$,
$\, \tilde{a} (\eta) \sim \tilde{a}_0 \eta^{\tilde D}$ with ${\tilde D} > 0$.
We then find that 
\be
R_H(\eta) = \eta ; \qquad
{V_H(\eta) \over \tilde{a}^{d-1}(\eta) A_H(\eta)} 
\sim {{\rm constant} \over \tilde{a}^{d-1}(\eta)}
{\int_0^{R_H} d \chi \sinh^{d-1} \chi \over \sinh^{d-1} R_H} \, .
\label{postbound}
\ee
For $\vert \gamma \vert \le 1/\sqrt{d}$,
the right-hand side of Eq.(\ref{postbound}) 
is a monotonic decreasing function
and again the holography is satisfied for all subsequent time 
provided it were at the Planck epoch. 
For $ \vert \gamma \vert \ge 1/\sqrt{d}$,  
the right-hand side of Eq.(\ref{postbound}) is 
a bounded function with 
a maximum  at some finite conformal time. 
For either cases, we see that the holography condition at the Planck epoch  
can be satisfied since the `graceful exit' connects the post-big bang phase 
to the holography compatible, pre-big bang phase.

Moreover, after the `graceful exit' is taken into account, the `effective 
particle horizon' $R^{\rm eff}_H (\eta) = \int_{-\infty}^\eta 
d \chi $
in conformal coordinate becomes infinite and puts the holography condition
most stringent. Nevertheless, as in the pre-big bang phase, the
ratio $V_H / A_H $ stays at a finite value $1/(d-1)$ and the holography can 
be easily satisfied.

Finally, several remarks are in order.
First, we have seen that the holography requires a smooth `branch change' 
interpolating the pre- and the post-big bang phases. During this epoch, unlike
adiabatic pre- or post-big bang phases, copious particle production is expected due to nonadiabatic change of the spacetime and the dilaton background.
The process thus produces particle entropy for the subsequent evolution of the
Universe. In this regard, Veneziano has informed us of his interesting 
result~\cite{veneziatoappear} indicating that the particle entropy is produced 
in such a way to saturate the holographic bound.   
Second, throughout the paper, 
we have stated the holographic principle in terms of
explicit solution of scale factor and dilaton. However, the statement of the
holography condition itself involves only {\sl global} aspects of the spacetime
(past light-cone, cosmological singularity etc.). This hints that the
cosmological holographic principle may be rephrased entirely in terms of
global data of the Universe and matter. It would be interesting to formulate
the principle along this direction explicitly.
Third, our analysis has been on the assumption of homogeneity and isotropy. 
It would
be interesting to see if the holography still works once such an assumption is
dropped, especially in light of possible non-genericity~\cite{buonanno}
of and recent criticism~\cite{turnerweinberg} to the homogeneous string
cosmology.

In this paper, we have tested the holographic principle to non-singular,
string inflationary cosmology. Emerging from our analyis is a picture that 
there exists a striking parallel between the comoving {\sl spatial} 
volume-to-area ratio in comoving coordinates for the open, flat and closed 
Universes and the Euclidean {\sl spacetime} volume-to-area ratio for the 
anti-de Sitter, Minkowski and de Sitter spaces respectively. 
For open Universe and anti-de Sitter spacetime, 
the respective ratio approaches a constant value and the holography
is satisfied. Likewise, for flat Universe and Minkowski spacetime, the 
respective ratio diverges linearly with the system size and are apt to violate 
the holography. For closed Universe and de Sitter spacetime, the area cannot 
be defined in an unambiguous manner and the holography is not subsistent at 
all. 

We are grateful to M. Gasperini and G. Veneziano for useful correspondences.

\end{document}